\documentclass[journal,twoside]{IEEEtran}

\ifCLASSINFOpdf
\usepackage[pdftex]{graphicx}

\else

\fi

\showoutput
\usepackage[cmex10]{amsmath}
\usepackage{amssymb}   % \triangleq 
\usepackage{amsxtra}
\usepackage{amscd}
\usepackage{amsthm}
\usepackage{textcomp}
\usepackage{graphicx}  
\usepackage{algorithm}
\usepackage{balance}
\usepackage{array}  
\usepackage{multirow}  
\usepackage{amsmath}
\usepackage{cite}
\usepackage{times}

\usepackage[noend]{algpseudocode}

\makeatletter
\def\BState{\State\hskip-\ALG@thistlm}

\makeatother
\usepackage{blindtext}
\usepackage{array}

\usepackage{url}
\usepackage{color,soul} % for highlights
\soulregister\cite7
\soulregister\ref7
\soulregister\pageref7
\begin{document}
	
	\title{{\huge Hybrid RIS-Empowered Reflection and Decode-and-Forward Relaying for Coverage Extension }}

	\author{Ibrahim~Yildirim,~\IEEEmembership{Graduate Student Member,~IEEE}, Fatih~Kilinc,~\IEEEmembership{Student Member,~IEEE},\\ Ertugrul~Basar,~\IEEEmembership{Senior Member,~IEEE} and George C. Alexandropoulos,~\IEEEmembership{Senior Member,~IEEE} \vspace*{-0.25cm}

	\thanks{I. Yildirim is with  Faculty of Electrical and Electronics Engineering, Istanbul Technical University and also with Koç University, Department of Electrical and Electronics Engineering, 34469 Istanbul, Turkey. (e-mail: yildirimib@itu.edu.tr)} 
	\thanks{F. Kilinc and E. Basar are with CoreLab, Koç University, Department of Electrical and Electronics Engineering, 34469 Istanbul, Turkey. (e-mail: fkilinc20@ku.edu.tr; ebasar@ku.edu.tr)}
	\thanks{G. C. Alexandropoulosis is with the Department of Informatics and Telecommunications, National and Kapodistrian University of Athens, 15784 Athens, Greece. (e-mail: alexandg@di.uoa.gr)} }
%	\author{\IEEEauthorblockN{Ibrahim Yildirim\textsuperscript{$\ast$,$\bullet$}, Fatih Kilinc\textsuperscript{$\ast$}, Ertugrul Basar\textsuperscript{$\ast$} and George C. Alexandropoulos\textsuperscript{$\odot$}}
%		\IEEEauthorblockA{\textsuperscript{$\ast$}CoreLab, Department of Electrical and Electronics Engineering, Ko\c{c} University, Sariyer 34450, Istanbul, Turkey  \\
%			\textsuperscript{$\bullet $}Faculty of Electrical and Electronics Engineering, Istanbul Technical University, Sariyer  34469, Istanbul, Turkey.  \\ 
%			\textsuperscript{$\odot $}Department of Informatics and Telecommunications, National and Kapodistrian University of Athens, Greece
%			\\
%			Email: yildirimib@itu.edu.tr, fkilinc20@ku.edu.tr, ebasar@ku.edu.tr}}

	\maketitle

	\begin{abstract}
%	In this letter, we introduce two hybrid reconfigurable intelligent surface (RIS) and relay-assisted transmission schemes by exploiting the most attractive advantages of RIS and relaying technologies.  The proposed hybrid schemes offer a flexible and cost-effective solution for possible disruptive effects encountered in the future wireless networks. Moreover, a maximized achievable rate is obtained for the hybrid systems by the proposed  sequential optimization algorithm. Our computer simulations and theoretical analysis  demonstrate that RIS and relaying technologies enhance the achievable rate and error performance remarkably by working as complementary to each other rather than being seen as an alternative to each other. 	
		In this letter, we introduce two hybrid transmission schemes combining a passive reconfigurable intelligent surface (RIS) with decode-and-forward relaying in a synergistic manner. The proposed schemes offer a flexible as well as cost- and power-efficient solution for coverage extension in future generation wireless networks. We present closed-form expressions for the end-to-end signal-to-noise ratio of both schemes and a sequential optimization algorithm for the power allocation and the RIS phase configurations. Our computer simulations and theoretical analysis demonstrate that the RIS and relaying technologies enhance the achievable rate and error performance remarkably when working complementary to each other, rather than being considered as competing technologies.
	\end{abstract}
	
	\begin{IEEEkeywords}
		6G, relaying, reconfigurable intelligent surface.\vspace{-0.3cm}
	\end{IEEEkeywords}

	%	\linenumbers

	\IEEEpeerreviewmaketitle
	
	\vspace*{-0.3cm}
	\section{Introduction}

	\IEEEPARstart{T}{he} unprecedented increase in data traffic in recent years has led researchers to explore new horizons beyond the existing paradigms. Although many innovative applications, such as virtual reality, autonomous vehicles, and telemedicine, are being introduced with the fifth-generation (5G) wireless networks, engineers and researchers have begun to establish the foundation for sixth-generation (6G) wireless networks to fulfill the stringent requirements of the upcoming Internet-of-Everything (IoE) era \cite{6G_roadmap}. In order to accomplish these requirements, reconfigurable intelligent surfaces (RISs) have been regarded as one of the key technologies for 6G wireless networks, owing to their capability to reconfigure the propagation environment with software-controlled reflection \cite{Basar_Access_2019,SimRIS_Mag,Holographic}. 
		
%	In order to accomplish these requirements, reconfigurable intelligent surfaces (RISs) have been regarded as one of the key components of future wireless networks by reconfiguring the propagation environment with software-controlled reflection \cite{Basar_Access_2019,SimRIS_Mag}. 
%	A typical RIS has a two-dimensional structure, which comprise large number of passive and low-cost reflectors embedded in the planar surface, and each of these reflectors is capable of smartly controlling the phase of the incoming signal.
	
	 RIS-empowered wireless transmission has recently drawn substantial attention from both academia and industry \cite{SimRIS_jour_new,DiRenzo_smart,yildirim2019propagation}. 
	 There are many research efforts examining the differences between RIS and relaying technologies. Although RISs and relaying are similar technologies used to alleviate the blockage effect and enhance the system performance, the received signal is actively processed at the relay by regenerating and retransmitting with an amplification, while the RIS only reflects the incident signal without any active transmit module \cite{Huang_2019}. In \cite{Emil_DF}, half-duplex decode-and-forward (DF) relaying and RIS-assisted transmission are compared in terms of energy efficiency, and it has been shown that the performance of a single antenna relay is achievable when hundreds of reflecting elements are used at the RIS. Moreover, a full-duplex relay-assisted transmission scheme comprising two horn antennas and two RISs very close to the relay has been proposed in \cite{AlkhateebRIS}, and considerable improvements in achievable rate  are achieved even with a low number of reflective elements. However, the consideration of two horn antennas for the full-duplex relay will cause high hardware cost, rendering end-to-end channel estimation difficult in the  case of large numbers of transmission nodes and terminals. Recently in \cite{Chambers_relay}, the authors proposed a hybrid transmission scenario  combining a DF relay with an RIS  and showed that using a single relay for low and moderate signal-to-noise ratio (SNR) values outperforms an RIS with massive number of reflecting elements, while for high SNRs, RIS-assisted	 transmissions are preferable.

As pointed out in the previous discussion, most of the studies related to relaying and RISs consider the one of these technologies as an alternative solution to the other. Against this background, in this study, we aim to find the best transmission scenarios where these two exciting technologies work in harmony with each other. The relay can be considered as an additional performance-enhancing component to the RIS-assisted transmission scenario and the hybrid transmission can be conducted by activating the relay to compensate for rapid deterioration in the channel quality.
  Our main motivation is to design hybrid RIS- and relay-aided systems with low operational complexity, while keeping the advantages of RIS and relaying technologies. To this end, two different cost efficient scheme are proposed for coverage extension by avoiding multi-hop relaying, and the best positioning for the RIS and the relay are investigated by considering that the direct communication link between the end terminals is blocked.\vspace*{-0.4cm}

\let\thefootnote\relax\footnote{\textit{Notation:} Boldface lowercase and capital letters, respectively, represent vectors and matrices. The transpose of $\mathbf{A}$ is represented by $\mathbf{A}^T$ and $\mathbb{C}$ is the complex number set. $\mathbb{E}[.]$ and  $|.|$ denote the expected and absolute values. $\mathcal{CN}(\mu,\sigma^2)$ denotes a complex Gaussian distributed random variable with $\mu$ mean and $\sigma^2$ variance.}

\section{System and Channel Models}
In this section, we introduce the proposed hybrid RIS and relay-assisted communication systems as well as provide their working principles. Two novel hybrid transmission concepts are proposed by considering that an RIS and a DF relay are placed between the source (S) and the destination (D). It is assumed that the relay operates in half-duplex mode, the RIS consists of $ N $ reflecting elements and the intelligent phase adjustment of the RIS is performed under perfect channel state information. Furthermore, all wireless channels are assumed to exhibit Rician fading with  a factor of $K$.

\vspace*{-0.3cm}
\subsection{Joint RIS and Relay Transmission Scheme}
\begin{figure}[!t]
	\begin{center} 
		\includegraphics[width=0.9\columnwidth]{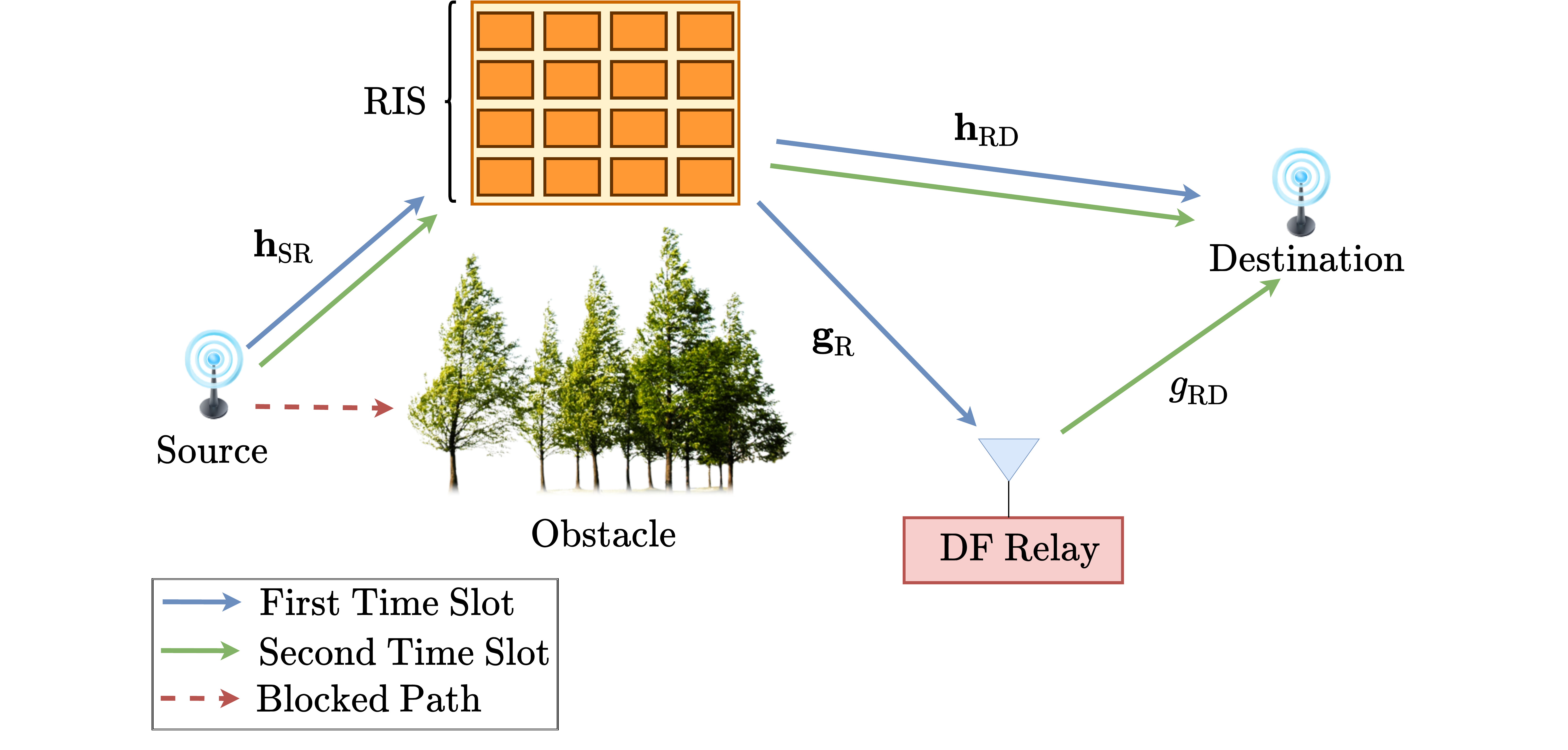}
		\vspace*{-0.3cm}\caption{The system model of the joint RIS and DF relay transmission scheme where the RIS and the relay are placed close to the S and D, respectively.}\vspace*{-0.5cm}
		\label{fig:Fig1}
	\end{center}\vspace*{-0.2cm}
\end{figure}

The system model of the joint RIS and relay transmission scheme is illustrated in Fig. \ref{fig:Fig1}. As seen from Fig. \ref{fig:Fig1},  S and  D are equipped with a single antenna and the channels between the S-D link and the S-relay link are not available due to the blockage. Since the signals at the millimeter and THz frequencies are highly vulnerable to the blockage, the reliability of transmission at high distances can be compromised. In this scenario, the coverage area of the communication is extended by placing an RIS in a position close to  S, while it is aimed to reach the signal via the DF relay placed close to  D.

In the first time-slot, the received signal at the relay and D are respectively given by  \vspace*{-0.1cm}
\begingroup\makeatletter\def\f@size{9}\check@mathfonts\begin{align}\label{eq:rec1_C1}
y_\text{R}=& \sqrt{P_1}\mathbf{g}_\text{R}^T\mathbf{\Phi}_1\mathbf{h}_\text{SR}s+n_\text{R}, \\
y_\text{D}^I=& \sqrt{P_1}\mathbf{h}_\text{RD}^T\mathbf{\Phi}_1\mathbf{h}_\text{SR}s+n_\text{D}^{I}
\end{align} \endgroup
where $ \mathbf{g}_\text{R}\in\mathbb{C}^{N\times1} $, $ \mathbf{h}_\text{SR}\in\mathbb{C}^{N\times1} $, and $\mathbf{h}_\text{RD}\in\mathbb{C}^{N\times1} $ are the fading channel coefficients between the RIS-relay, S-RIS, and RIS-D links, $ P_1 $ is the transmission power of S, $ \mathbf{\Phi}_1 =\text{diag}\left\lbrace e^{j\phi_1^1},\dots, e^{j\phi_1^N} \right\rbrace $, with $ \phi_1^i$ being the phase shift introduced by $i$th element of the RIS in the first time slot, $ s $ is the transmitted signal from  S with unit-average-power ($\mathbb{E}[|s|^2]=1$), $ n_\text{R} $ and $ n_\text{D}^{I} $ are the additive white Gaussian noise (AWGN) terms at the relay and  D, which are distributed according to  $\mathcal{CN}(0,N_0) $.

Assuming that the DF relay has a perfect decoding capability, the received signal at D in the second time slot can be expressed as follows:\vspace*{-0.2cm}
\begingroup\makeatletter\def\f@size{9}\check@mathfonts\begin{align}\label{eq:rec2_C2}
y_\text{D}^{II}= \sqrt{P_2}\mathbf{h}_\text{RD}^T\mathbf{\Phi}_2\mathbf{h}_\text{SR}s+\sqrt{P_3}{g}_\text{RD}s+n_\text{D}^{II}
\end{align} \endgroup
where $ {g}_\text{RD} \in\mathbb{C} $ is the fading channel coefficient between the relay-D link, $ P_2 $ and $ P_3 $ are respectively the transmission power of S and the relay in the second time slot, $ \mathbf{\Phi}_2 =\text{diag}\left\lbrace e^{j\phi_2^1},\dots, e^{j\phi_2^N} \right\rbrace $, with $ \phi_2^i$ being the phase shift of $i$th element of the RIS in the second time slot and  $ n_\text{D}^{II} $ is the AWGN term distributed according to $\mathcal{CN}(0,N_0) $. In \eqref{eq:rec2_C2}, the first term is denotes the RIS-based signal, while the second term represents the signal transmitted by the relay after perfectly decoded as $s$.

By considering intelligent phase shifting capability of the RIS, the maximized received SNR at D in the first time slot is expressed by \vspace*{-0.2cm}
\begingroup\makeatletter\def\f@size{8.5}\check@mathfonts\begin{align} { \label{eq:SNR1_C1}
\gamma_1=\max_{\mathbf{\Phi}_1} \dfrac{P_1\left| \mathbf{h}_\text{RD}^T\mathbf{\Phi}_1\mathbf{h}_\text{SR}\right|^2 }{N_0}  = \dfrac{P_1\left| \sum\limits_{i=1}^{N}|{h}_\text{RD,i}||{h}_\text{SR,i}|\right|^2 }{N_0}= \dfrac{P_1A^2 }{N_0}}
\end{align} \endgroup
where $ A $ has the Gaussian distribution for $N \gg 1$ due to central limit theorem (CLT) with the following mean and variance \cite{yildirim2019propagation}:\vspace*{-0.2cm}
\begingroup\makeatletter\def\f@size{8.5}\check@mathfonts\begin{align}\label{eq:C1_par1}
\mu_A =&\ \dfrac{N\sqrt{P_L^{R_1}}\pi }{4(K+1)}L_{1/2}^2(-K), \nonumber \\
\sigma_A^2 =& NP_L^{R_1}- \dfrac{N{P_L^{R_1}}\pi^2 }{16(K+1)^2}L_{1/2}^4(-K)
\end{align} \endgroup
where $P_L^{R_1}$ is the path loss for the S-D link and $ L_{n}(x)=\frac{e^x}{n!}\frac{d^n}{dx^n}(e^{-x}x^n) $ is the Laguerre polynomials of degree $ n $.
\normalsize Since $\gamma_1$ is a non-central chi-square distributed random variable with one degree of freedom, its moment generating function (MGF) \cite{Proakis} is calculated by\normalsize
\begingroup\makeatletter\def\f@size{8}\check@mathfonts\begin{align}\label{eq:MGF1_C1} 
&M_{\gamma_1}(s)=\left({1-\dfrac{sNP_L^{R_1}\left[ 16(K+1)^2-\pi^2L_{1/2}^4(-K)\right]P_1 }{8(K+1)^2N_0} }\right)^{-1/2}  \nonumber \\
&\times\exp\left( \dfrac{\dfrac{sP_L^{R_1}N^2\pi^2L_{1/2}^4(-K)P_1}{16(K+1)^2N_0}}{1-\dfrac{sNP_L^{R_1}\left[ 16(K+1)^2-\pi^2L_{1/2}^4(-K)\right]P_1 }{8(K+1)^2N_0}}\right). 
\end{align} \endgroup

%\footnotesize\begin{align}\label{eq:5} 
%M_{\gamma_1}(s)=\left(\dfrac{1}{1-\dfrac{sNP_L^{R_1}\left[ 16(K+1)^2-\pi^2L_{1/2}^4(-K)\right]P_1 }{8(K+1)^2N_0} }\right)^{1/2}  \nonumber \\
%\times\exp\left( \dfrac{\dfrac{sP_L^{R_1}N^2\pi^2L_{1/2}^4(-K)P_1}{16(K+1)^2N_0}}{1-\dfrac{sNP_L^{R_1}\left[ 16(K+1)^2-\pi^2L_{1/2}^4(-K)\right]P_1 }{8(K+1)^2N_0}}\right) 
%\end{align} \normalsize

By considering the similar phase adjustment in the first time slot and assuming $P_2=P_3$ due to simplicity, the maximized received SNR at the D in the second time slot is obtained as
\begingroup\makeatletter\def\f@size{8}\check@mathfonts\begin{align}\label{eq:SNR2_C1}
\gamma_2=&\max_{\mathbf{\Phi}_2} \dfrac{P_2\left|g_{RD}+ \mathbf{h}_\text{RD}^T\mathbf{\Phi}_2\mathbf{h}_\text{SR}\right|^2 }{N_0} \nonumber \\=&  \dfrac{P_2\left| |g_{RD}|+ \sum\nolimits_{i=1}^{N}|{h}_\text{RD,i}||{h}_\text{SR,i}|\right|^2 }{N_0}= \dfrac{P_2B^2 }{N_0}.
\end{align} \endgroup
Here, due to the Lyapunov variant of the CLT \cite{Lyapunov}, $B$ is a Gaussian distributed random variable for $N\gg 1$ with the following mean and variance respectively
\begingroup\makeatletter\def\f@size{7.9}\check@mathfonts\begin{align}\label{eq:C1_par2}
\mu_B=&P_L^{D} \sqrt{\dfrac{\pi}{4(K+1)}}L_{1/2}(-K)+\dfrac{N\sqrt{P_L^{R_1}}\pi }{4(K+1)}L_{1/2}^2(-K), \nonumber \\
\sigma_B^2=& P_L^{D}-\dfrac{P_L^{D} \pi L_{1/2}^2(-K)}{(K+1)}+NP_L^{R_1} - \dfrac{N{P_L^{R_1}}\pi^2 L_{1/2}^4(-K)}{16(K+1)^2}
\end{align} \endgroup
where $P_L^{D}$ is the path loss for the relay-D link. Similar to $\gamma_1$, $\gamma_2$ follows the non-central chi-square distribution with one degree of freedom and its MGF is calculated by
%\begingroup\makeatletter\def\f@size{8}\check@mathfonts\begin{align}\label{eq:MGF2_C1} 
%M_{\gamma_2}(s)=\left( \dfrac{1}{1-\dfrac{2s\sigma_B^2P_2}{N_0}}\right)^{1/2}\exp\left( \dfrac{\dfrac{s \mu_B^2P_2}{N0}}{1-\dfrac{2s\sigma_B^2P_2}{N_0}}\right)  . 
%\end{align} \endgroup 
\begingroup\makeatletter\def\f@size{8}\check@mathfonts\begin{align}\label{eq:MGF2_C1} 
M_{\gamma_2}(s)=\left(1-\dfrac{2s\sigma_B^2P_2}{N_0}\right)^{-1/2}\exp\left( \dfrac{s \mu_B^2P_2/N_0}{1-2s\sigma_B^2P_2/N_0}\right)  . 
\end{align} \endgroup \vspace*{-0.4cm}

%\begin{figure*}[!t]		
%	{\footnotesize\begin{align} \label{eq:8}
%		M_{\gamma_2}(s)=\left({1-\dfrac{s\left[ 16(K+1)^2(NP_L^{R_1}+P_L^D)-NP_L^{R_1}\pi^2L_{1/2}^4(-K)-P_L^D4(K+1)\pi L_{1/2}^2(-k)\right]P_2 }{8(K+1)^2N_0} }\right)^{-1/2} 
%		\nonumber \\ \times\exp\left( \dfrac{\dfrac{s\left( NP_L^{R_1}\pi L_{1/2}^2(-K)+\sqrt{P_L^D4(K+1)\pi} L_{1/2}(-K) \right)^2 P_2}{16(K+1)^2N_0}}{1-\dfrac{s\left[ 16(K+1)^2(NP_L^{R_1}+P_L^D)-NP_L^{R_1}\pi^2L_{1/2}^4(-K)-P_L^D4(K+1)\pi L_{1/2}^2(-k)\right]P_2 }{8(K+1)^2N_0}}\right). 
%		\end{align}}
%	\hrulefill
%	\vspace*{-0.5cm}
%\end{figure*}

%\footnotesize\begin{align}\label{eq:8} 
%M_{\gamma_2}(s)=\left(\dfrac{1}{1-\dfrac{s\left[ 16(K+1)^2(NP_L^{R_1}+P_L^D)-NP_L^{R_1}\pi^2L_{1/2}^4(-K)-P_L^D4(K+1)\pi L_{1/2}^2(-k)\right]P_1 }{8(K+1)^2N_0} }\right)^{1/2}  \nonumber \\
%\times\exp\left( \dfrac{\dfrac{sP_L^{R_1}\pi^2L_{1/2}^4(-K)P_1}{16(K+1)^2N_0}}{1-\dfrac{sNP_L^{R_1}\left[ 16(K+1)^2-\pi^2L_{1/2}^4(-K)\right]P_2 }{8(K+1)^2N_0}}\right) 
%\end{align} \normalsize

\subsection{Integrated RIS and Relay Transmission Scheme}

In this hybrid scheme, we propose a transmission scheme where the RIS and the relay are integrated in the same device and positioned in between S and D as shown in Fig. 2. In contrast to the joint RIS and relaying scenario, we assume that S's signal reaches the  RIS-relay device, which is both reflected and DF relayed to D. We again assume here that the distance between S and D is considerably large, hence, the S-D direct communication link is blocked. 

%This scenario is designed to increase system performance by providing an additional transmission path, and can be considered as an alternative to joint RIS and relaying scenario, when the distance between S and D is considerably large and the S-D direct link is blocked.

In the first time-slot of the integrated RIS and relay transmission scheme, the received signal at the relay and D are respectively given by \vspace*{-0.4cm}
\begingroup\makeatletter\def\f@size{9}\check@mathfonts\begin{align}\label{eq:10}
y_\text{R}=& \sqrt{P_1}g_\text{SR}s+n_\text{R}, \\
y_\text{D}^I=& \sqrt{P_1}\mathbf{h}_\text{RD}^T\mathbf{\Phi}_1\mathbf{h}_\text{SR}s+n_\text{D}^{I},
\end{align} \endgroup
where $ g_\text{SR}\in \mathbb{C}$, $\mathbf{h}_\text{SR}\in \mathbb{C}^{N\times1}$ and $ \mathbf{h}_\text{RD} \in \mathbb{C}^{N\times1}$ are the fading channel coefficients between the S-relay, S-RIS and RIS-D links, $ P_1 $ is the transmission power of S, $ s $ is the transmitted signal from S with the unit-average-power, $ n_\text{R} $ and $n_\text{D}^{I}$ are the AWGN at the relay and D, which are distributed according to $ \mathcal{CN}(0,N_0) $.

\begin{figure}[!t]
	\begin{center}
		\includegraphics[width=0.9\columnwidth]{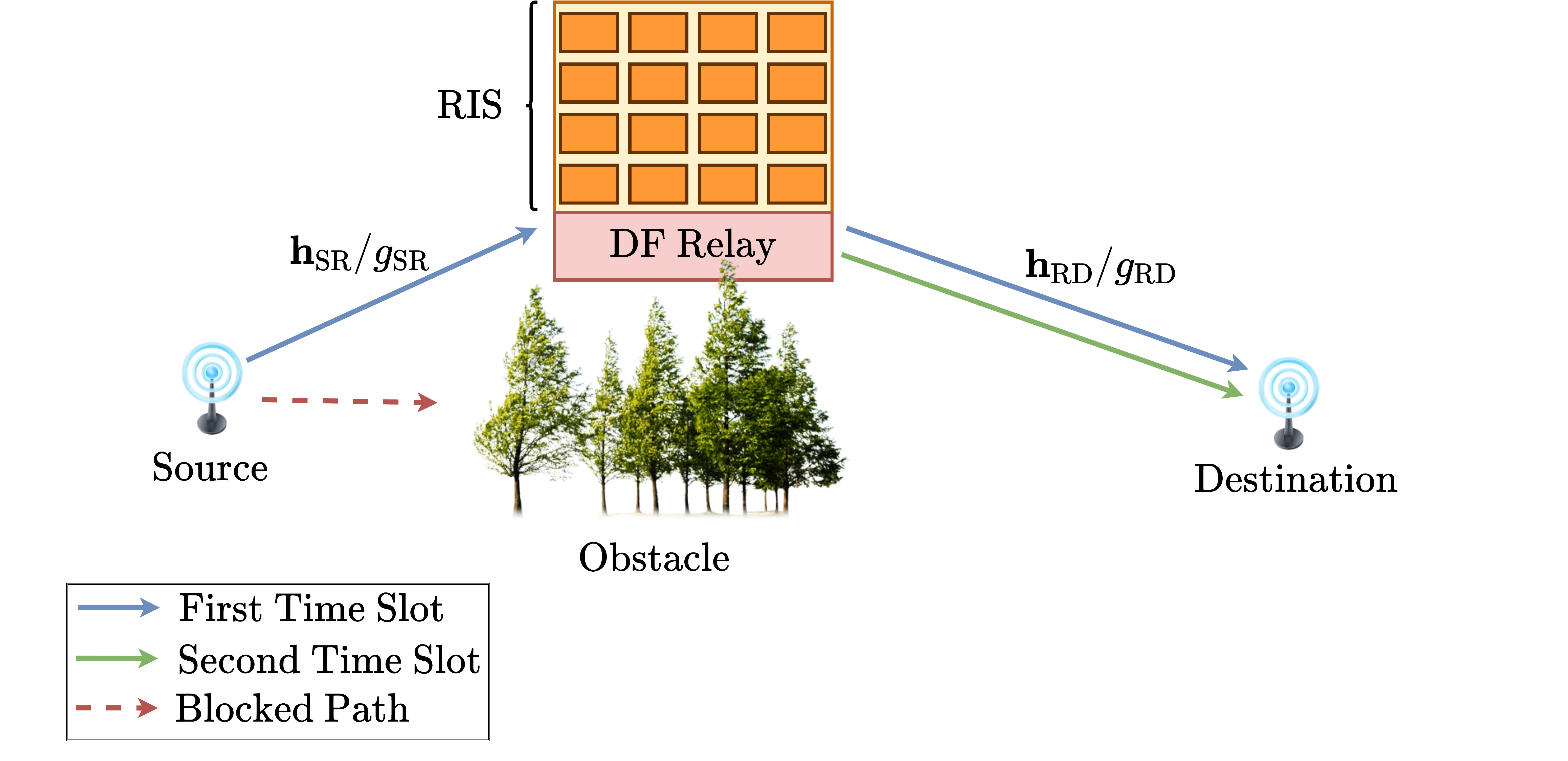}
		\vspace*{-0.3cm}\caption{The system model of the integrated RIS and DF relay transmission scheme.}\vspace*{-0.5cm}
		\label{fig:Fig2}
	\end{center} \vspace*{-0.2cm}
\end{figure}

The received signal transmitted through the perfect DF relay in the second time slot is obtained as \vspace*{-0.1cm}
\begingroup\makeatletter\def\f@size{9}\check@mathfonts\begin{align}\label{eq:11}
y_\text{D}^{II}=\sqrt{P_2}{g}_\text{RD}s+n_\text{D}^{II},
\end{align} \endgroup 
where $ {g}_\text{RD} $ is the fading channel coefficients between the relay-D link, $ P_2  $is the transmission power of relay in the second time slot, and $ n_\text{D}^{II} $ is the AWGN as $ n\sim\mathcal{CN}(0,N_0) $.

Similar to the joint RIS and relay case, the received SNR in the first time slot is expressed as in \eqref{eq:SNR1_C1}. 
%\vspace*{-0.2cm}
%\begingroup\makeatletter\def\f@size{9}\check@mathfonts\begin{align}\label{eq:12}
%\gamma_1=\max_{\mathbf{\Phi}_1} \dfrac{P_1\left| \mathbf{h}_\text{RD}^T\mathbf{\Phi}_1\mathbf{h}_\text{SR}\right|^2 }{N_0} = \dfrac{P_1\left| \sum\limits_{i=1}^{N}|{h}_\text{RD,i}||{h}_\text{SR,i}|\right|^2 }{N_0}= \dfrac{P_1A^2 }{N_0}
%\end{align} \endgroup
%where $A$ follows Gaussian distribution has mean and variance as in \eqref{eq:C1_par1}. 
Therefore, MGF of $\gamma_1$ is obtained as in \eqref{eq:MGF1_C1} for integrated RIS and relay transmission. Furthermore, the received SNR at the D in the second time slot is given by $ \gamma_2=P_2\left| {g}_\text{RD}\right|^2/N_0 $
%\begin{align}\label{eq:13}
%\gamma_2=\dfrac{P_2\left| {g}_\text{RD}\right|^2 }{N_0}
%\end{align} 
where $\gamma_2$ has non-central chi-square distribution with two degrees of freedom and its MGF is expressed as
%\begingroup\makeatletter\def\f@size{8}\check@mathfonts\begin{align}\label{eq:14}
%M_{\gamma_2}(s)=\left( \dfrac{1}{1-\dfrac{P_2 P_L^D}{2(K+1)N_0}}\right)^{1/2}\exp\left(\dfrac{\dfrac{sP_2P_L^DK}{(K+1)N_0}}{1-\dfrac{P_2P_L^D}{2(K+1)N_0}} \right)  . 
%\end{align}\endgroup
\begingroup\makeatletter\def\f@size{8}\check@mathfonts\begin{align}\label{eq:14}
M_{\gamma_2}(s)=\left(1-\dfrac{P_2 P_L^D}{2(K+1)N_0}\right)^{-1/2}\exp\left(\dfrac{2sP_2P_L^DK}{2(K+1)N_0-P_2P_L^D}\right)  . 
\end{align}\endgroup

\vspace*{-0.2cm}
\section{Performance Analysis and Sequential Optimization}
In this section, theoretical average symbol error probability (SEP) and achievable rate expressions of the  joint and integrated RIS and relay schemes  are evaluated under the ideal DF relay assumption. Further, a sequential optimization algorithm is proposed to maximize the achievable rate of these hybrid schemes under non-ideal relay assumption. By considering the maximum likelihood detection rule, the received signals can be combined by using the maximal ratio combining (MRC) technique for both schemes, and the total SNR at D can be obtained as $\gamma_\text{tot}=\gamma_1+\gamma_2$. Therefore, the MGF of the total SNR is calculated by $M_{\gamma_\text{tot}}(s)=M_{\gamma_1} (s)M_{\gamma_2}(s)$. Thus, the average (SEP) of the proposed schemes for  $M$-phase shift keying (PSK) signaling (as a representative example) is obtained as \cite{Simon} 
\begingroup\makeatletter\def\f@size{9}\check@mathfonts
 \begin{align}\label{eq:SEP}
P_e = \frac{1}{\pi}\int_0^{(M-1)\pi/M} M_{\gamma_{\text{tot}}} \left(-\frac{\sin(\pi/M)}{\sin^2\eta}\right)d\eta.
\end{align}\endgroup
For binary PSK (BPSK) case, \eqref{eq:SEP} can be easily modified for $M=2$. 
%\begingroup\makeatletter\def\f@size{9}\check@mathfonts\begin{align}\label{eq:SEP2}
%P_e = \frac{1}{\pi}\int_0^{\pi/2} M_{\gamma_{\text{tot}}} \left(-\frac{1}{\sin^2\eta}\right)d\eta.
%\end{align}\endgroup
Further, under the perfect DF relay assumption, the achievable rate for both transmission scheme is calculated by $R_\text{Hybrid} = \log_2\left( 1+\gamma_{\text{tot}}\right)$.
%\begingroup\makeatletter\def\f@size{9}\check@mathfonts\begin{align}\label{eq:Rate}
%R_\text{Hybrid} = \log_2\left( 1+\gamma_{\text{tot}}\right)= \log_2\left( 1+\left( \gamma_{1}+\gamma_{2}\right) \right).
%\end{align}\endgroup

On the other hand, we analyze the achievable rate performance of our hybrid RIS and relay transmission schemes, and propose the sequential phase and power optimization algorithm in order to maximize the achievable rate under the non-ideal DF relay assumption. The decoding performance of the transmitted symbol  degrades as the distance between the S and the relay increases. In this case, the transmit power may be adjusted to ensure that the maximized performance metrics are obtained. 
%The proposed concepts introduce RIS into the equation that enables the first time slot to be considered in SNR calculations which significantly improves the received SNR for large $  N $. The total power is distributed between first and second time slots, in a way, maximum channel capacity is achieved. 
Under the non-ideal DF relay, the achievable rate is calculated by
\begingroup\makeatletter\def\f@size{9}\check@mathfonts\begin{align}
R_\text{Hybrid}^i =\frac{1}{2}\log_2{\left(  1+\min\left\lbrace \gamma_{r}^i,\gamma_{tot}\right\rbrace\right) } , ~i\in\left\lbrace1,2 \right\rbrace 
\end{align}\endgroup
where $ \gamma_{r}^1 $ and $\gamma_{r}^2$ is the received SNR at the relay for joint ($i=1$) and integrated ($i=2$) RIS and relay case. 
 In order to achieve maximum achievable rate for the hybrid RIS and relay configuration in the joint RIS and relay transmission, the corresponding power optimization problem is formulated as
\begingroup\makeatletter\def\f@size{9}\check@mathfonts\begin{align}\label{eq:opt1}
R_\text{Hybrid}^1 =\max_{P_1} \quad & \min\left( \frac{P_1|\mathbf{g}_\text{R}^T\mathbf{\Phi_1}\mathbf{h}_\text{SR}|^2}{N_0},
\dfrac{P_2\text{A}^2 }{N_0}
+\dfrac{P_2\text{B}^2 }{N_0} \right)   \\
&\hspace*{0.1cm}\text{s.t.}\hspace*{0.5cm} P_2=(P_{\text{tot}}-P_1)/2,\quad P_1<P_{\text{tot}} \nonumber
\end{align}\endgroup
Here, we assumed that $P_2=P_3$ to have a convex optimization problem. Thus, the power constraint in \eqref{eq:opt1} is set by considering $P_t=P_1+2P_2$. The corresponding optimization problem for integrated RIS and relay case at D in the second time slot is expressed as \vspace*{-0.2cm}

\scriptsize\begin{algorithm}[!t]
	\caption{Sequential Phase and Power Optimization}\label{IniA}
	\begin{algorithmic}[1]
		\scriptsize
		\If{Joint RIS and relay transmission scheme}.
		\State $\textbf{Input:}\; N_0,P_t, \textbf{h}_\text{SR},\textbf{h}_\text{RD},\textbf{g}_\text{R},{g}_\text{RD}$.
		\State Set $P_2=(P_t-P_1)/2$.
		\State Optimize $\mathbf{\Phi_1}$ and $\mathbf{\Phi_2}$ by yielding the largest value of (4) and (7), respectively.
		\State Obtain $A,B$ by using input channels, $\mathbf{\Phi_1}$ and $\mathbf{\Phi_2}$.
		\State Find optimal solution to $P_1$ and $P_2$ via CVX yielding the largest value of (16)
		\State \textbf{return} $P_1,\mathbf{\Phi_1},\mathbf{\Phi_2}$.
		\Else $\;\text{(Integrated RIS and relay transmission scheme)}$
		\State $\textbf{Input:}\;N_0,P_t, \textbf{h}_\text{SR},\textbf{h}_\text{RD},{g}_\text{SR},{g}_\text{RD}$.
		\State Set $P_2=P_t-P_1$.
		\State Optimize $\boldsymbol{\Phi_1}$ by yielding the largest value of (13).
		\State Obtain $A$ by using input channels and $\mathbf{\Phi_1}$.
		\State Find optimal solution to $P_1$ and $P_2$ via CVX yielding the largest value of (17)
		\State \textbf{return} $P_1,\mathbf{\Phi_1}$
		\EndIf
	\end{algorithmic}
\end{algorithm}\vspace{-0.2cm}\normalsize 

%The corresponding optimization problem for Concept 2 is expressed as at the D in the second time slot is obtained by \vspace*{-0.2cm}
\begingroup\makeatletter\def\f@size{9}\check@mathfonts\begin{align}\label{eq:opt2}
R_\text{Hybrid}^2 =\max_{P_1} \quad & \min\left( \frac{P_1|{g}_\text{SR}|^2}{N_0},
\dfrac{P_2\text{A}^2 }{N_0}
+\dfrac{P_2\left| {g}_\text{RD}\right|^2 }{N_0} \right)   \\
&\hspace*{0.1cm}\text{s.t.}\hspace*{0.5cm} P_2=P_{\text{tot}}-P_1, \quad P_1<P_{\text{tot}} \nonumber
\end{align}\endgroup
Since S does not transmit in both time slots for the integrated RIS and relay scheme, the power constraint in \eqref{eq:opt2} is set by considering $P_t=P_1+P_2$.
Here, the total power  should be allocated between first and second time slots, in order to maximize the achievable rate. To this end, sequential phase and power optimization algorithm is proposed to sequentially optimize the phase responses of the RIS elements and transmit powers in order to obtain the maximized achievable rate. In this algorithm, the inputs are selected according to corresponding concept. First, $\mathbf{\Phi_1}$ and $\mathbf{\Phi_2}$ are optimized to maximize the received SNR in both time slots. Then, using this phase matrices, end-to-end channel expressions are obtained to modify SNR expressions. In the last step, the optimal value of $P_1$ is calculated by using CVX tool \cite{CVX} to achieve maximum rate. The overall steps of the sequential algorithm to solve (16) and (17) are summarized in Algorithm 1. \vspace*{-0.2cm}

\begin{figure}[!t]
	\begin{center}
		\includegraphics[width=0.85\columnwidth]{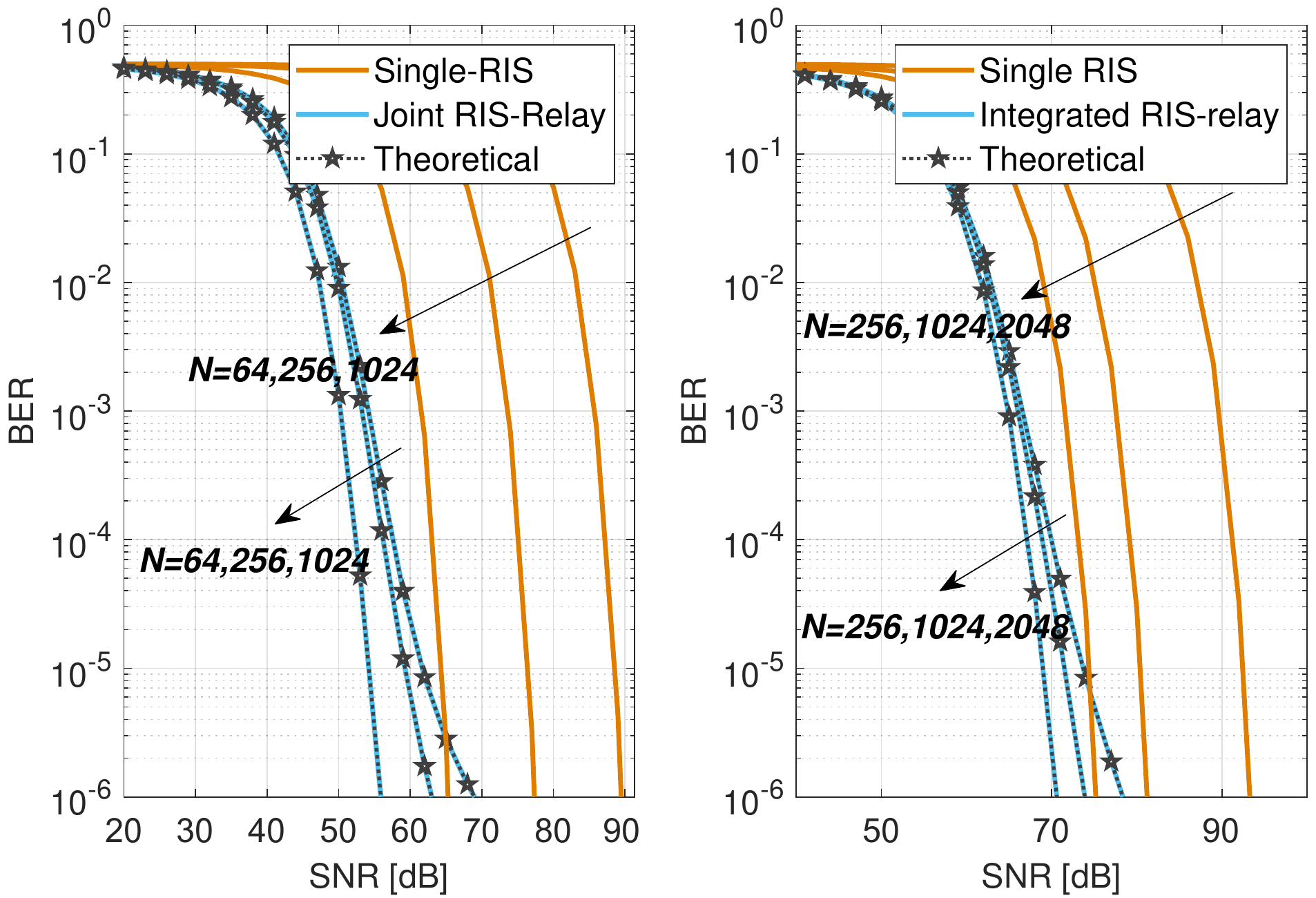}
		\vspace*{-0.3cm}\caption{BER performance of the (a) joint, and (b) integrated RIS and relay transmission schemes for increasing $N$ values.}\vspace*{-0.6cm}
		\label{fig:BER}
	\end{center}\vspace*{-0.2cm}
\end{figure}

\section{Numerical Results}

In this section, we illustrate the error and achievable rate performances of the proposed hybrid schemes via numerical simulations by considering path loss as well as channel fading. Rician factor is assumed as $K=10$ dB for all simulations, and path loss is modeled using the 3GPP Urban Micro (UMi) \cite{3GPP_phy2} and 5G UMi-Street Canyon \cite{5G_pathloss} path loss models for $ 2.4 $ GHz and $28$ GHz, respectively. Here, we consider a two-dimensional (2D) coordinate system for the positioning of the terminals, which are located in the $xy$-plane.

Figs. \ref{fig:BER}(a) and (b) demonstrate the theoretical and simulated average bit error rate (BER) performances of the proposed hybrid schemes with respect to SNR, which is defined as $P_{\text{tot}}/N_0$, for BPSK signaling and $P_1=P_2=5$ W at $ 2.4 $ GHz under increasing $ N $. In this comparison, a single RIS-assisted transmission scheme is considered as a benchmark system. As clearly seen from Figs. \ref{fig:BER}(a) and (b), the theoretical and simulation results are in perfect agreement for both scenarios. In Fig. \ref{fig:BER}(a), the simulations are conducted for the joint RIS and relay case when coordinates of the S, D, RIS, and the relay in the $xy$-plane are given as $ (5,0) $, $ (5,10) $, $ (0,15) $, and $ (10,35) $, respectively. Meanwhile, 2D coordinates of the S, D, and the RIS/relay are respectively given as $ (40,0) $, $ (40,75) $, and $ (0,35) $ for the integrated RIS and relay transmission in Fig. \ref{fig:BER}(b). For both scenarios, increasing $ N $ provides a significant improvement in BER over a single RIS-assisted transmission, and doubling $ N $ leads to about $ 5 $ dB gain in SNR for joint RIS and relay case, while it brings approximately $ 3 $ dB gain for integrated RIS and relay case. Since the effect of the relay in transmission diminishes for large $N$ values, the BER performances of the single-RIS and hybrid schemes get closer to each other. 
 It demonstrates that joint RIS and relay transmission scheme appears to be more efficient than the integrated RIS and relay scheme in terms of BER performance.

\begin{figure}[!t]
	\begin{center}
		\includegraphics[width=0.88\columnwidth]{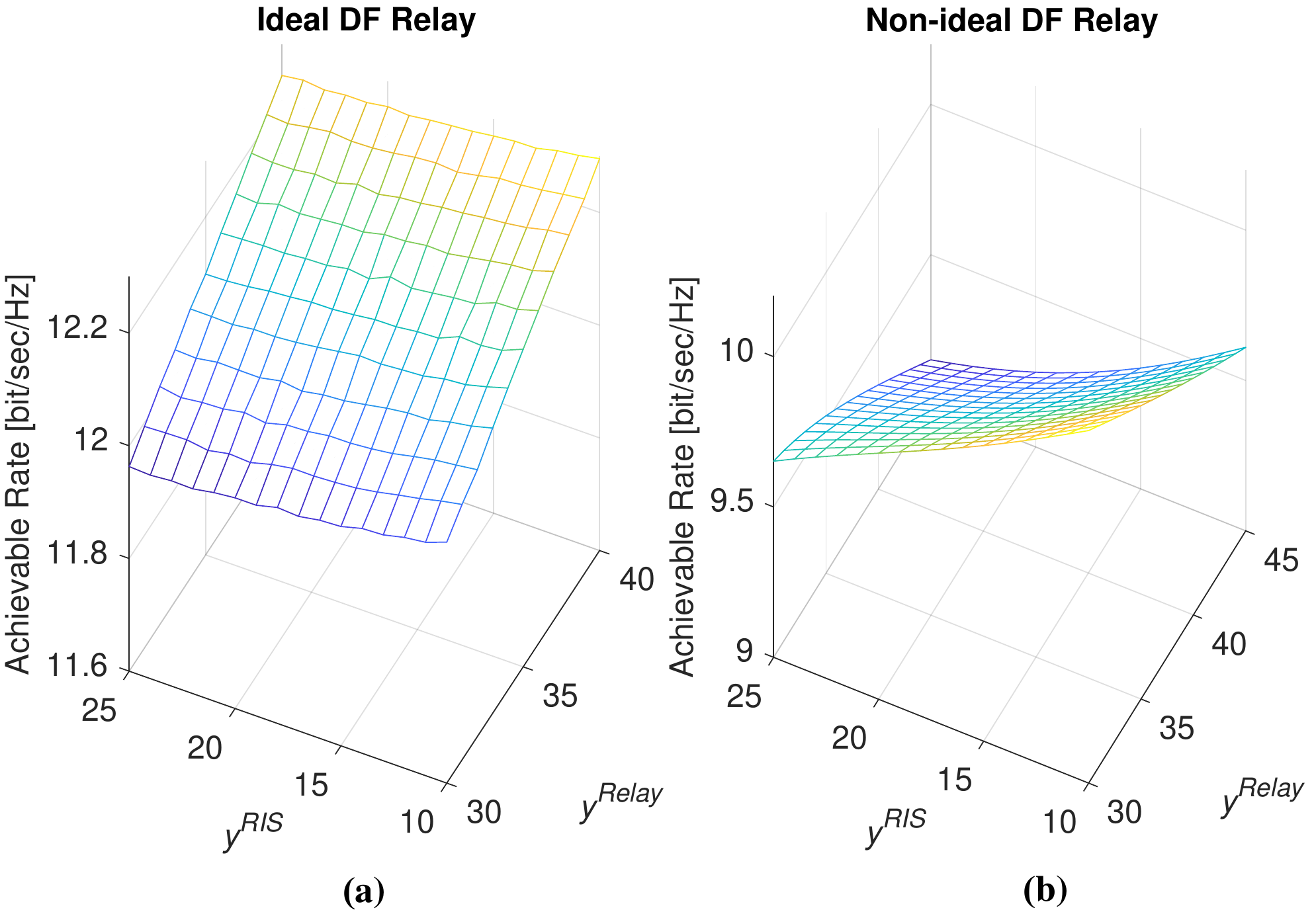}
		\vspace*{-0.3cm}\caption{Achievable rate  of the joint RIS and relay scheme when (a) ideal and (b) non-ideal relaying are considered for a varying distance in the $y$-axis between the RIS and the DF relay.}\vspace*{-0.6cm}
		\label{fig:3Dnon_id}
	\end{center}
\end{figure}
\begin{figure}[!t]
	\begin{center}
		\includegraphics[width=0.8\columnwidth]{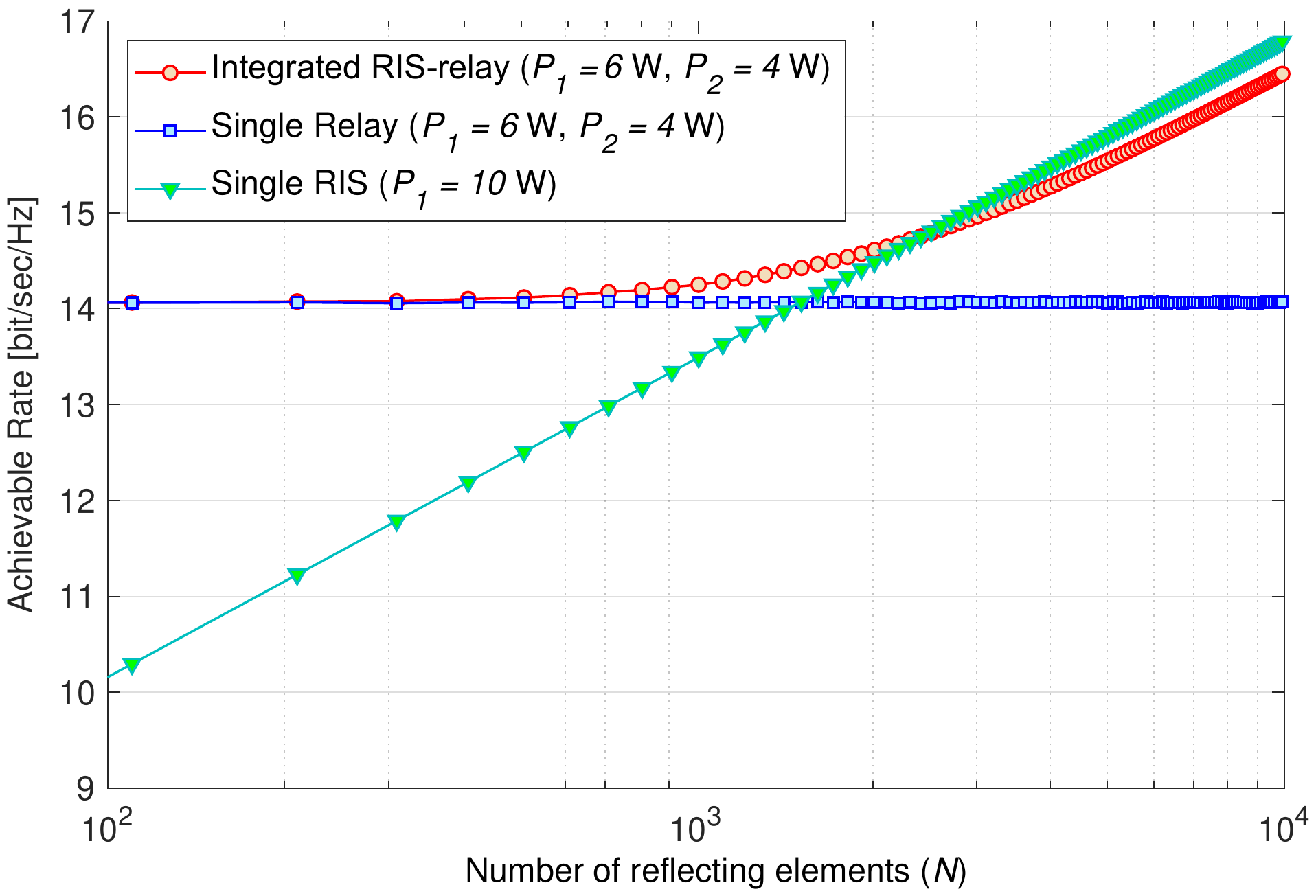}
		\vspace*{-0.3cm}\caption{Achievable rates with the two proposed RIS and DF relay schemes for varying $N$ values.}\vspace*{-0.6cm}
		\label{fig:rate_C2}
	\end{center}
\end{figure}

In Figs. \ref{fig:3Dnon_id}(a) and (b), the effect of varying RIS and relay positions in the $y$-axis on the achievable rate performance of the joint RIS and relay transmission scheme is investigated under perfect and imperfect DF relay assumption for $P_1=5$ W, $P_2=2.5$ W, $N_0=-130$ dBm and $N=2048$ at $28$ GHz. Here, the S, D, RIS, and the relay are located at $(10,0)$, $(10,60)$, $(0,y^{\text{RIS}})$ and $(25,y^{\text{Relay}})$ in the $xy$-plane, respectively. As seen from Fig. \ref{fig:3Dnon_id}(a), the best achievable rate performance is obtained when the RIS is close to  S and the relay is close to  D, while the proximity of the ideal relay to D is a more dominant factor in this scheme. On the other hand, as seen in Fig. \ref{fig:3Dnon_id}(b), since the RIS plays a more effective role in transmission when the relay does not have perfect decoding capability, the best achievable rate performance is obtained when the RIS is close to the S and the relay close to the RIS. In other words, the existence of a reliable channel between the RIS-relay will play a key role in achievable rate  under the non-ideal relay.

In Fig. \ref{fig:rate_C2}, the achievable rate  of the integrated RIS and relay transmission scheme  is compared to the single-RIS and single-relay assisted transmission scheme under varying $N$ at $2.4$ GHz. Here, the S, D, and the RIS/relay are located at $(20,0)$, $(20,55)$, and $(0,20)$, respectively. As shown, increasing $ N $ significantly improves the effect of the RIS in transmission both in the case of the single-RIS and the integrated RIS and relay scheme. Moreover, for the  latter scheme, these three cases can be also considered as three different operating modes, and the activation of the RIS and/or relay for transmission can be dynamically determined according to the requirements. In order to increase energy efficiency, if the number of reflecting elements of RIS is enough for a reliable communication, transmission can be conducted over a single RIS, while the hybrid transmission can be conducted by activating the relay to compensate rapid deterioration in the channel quality.

\begin{figure}[!t]
	\begin{center}
		\includegraphics[width=0.85\columnwidth]{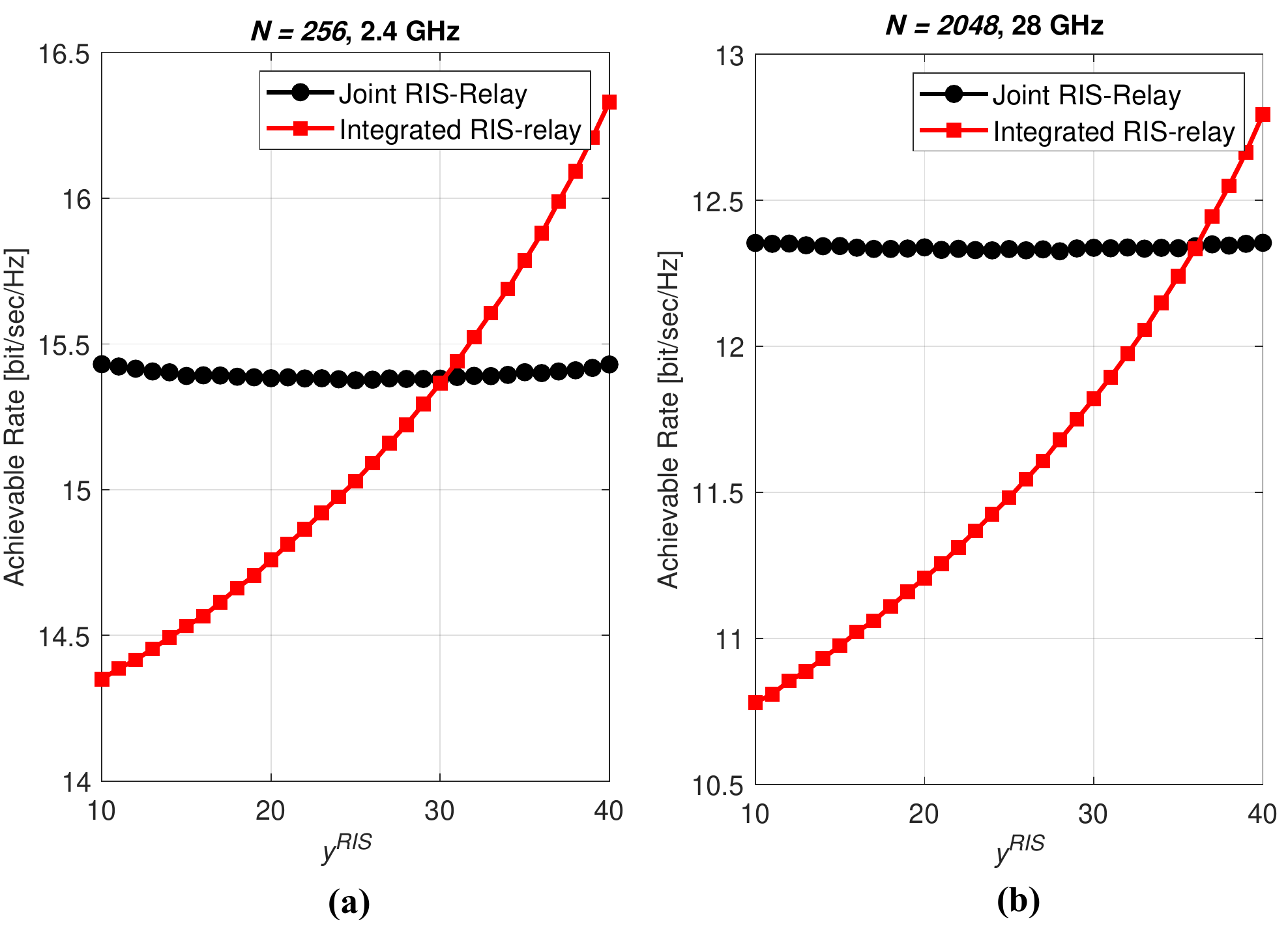}
		\vspace*{-0.3cm}\caption{Achievable rates of the joint and integrated RIS and relay transmission schemes for (a) $N=256$ at $2.4$ GHz and (b) $N=2048$ at $28$ GHz.}\vspace*{-0.6cm}
		\label{fig:rate_C1vsC2}
	\end{center}
\end{figure}

Figs. \ref{fig:rate_C1vsC2}(a) and (b) exhibit the achievable rate comparison of the joint RIS and relay transmission scheme for $ P_1=5 $ W and $ P_2=P_3=2.5 $ W, and the integrated RIS and relay transmission scheme for $ P_1=P_2=5 $ W, under changing the RIS position through the $y$-axis. Here, the S,  D, and RIS are respectively located at $(5,0)$, $(5,50)$, and $(0, y^{\text{RIS}})$ for both schemes, while the relay is located at $(10,35)$ for the integrated RIS and relay case. The results shown in Figs. \ref{fig:rate_C1vsC2}(a) and (b) indicate that joint RIS and relay scheme outperforms integrated RIS and relay transmission scheme, although the advantage of  the joint RIS and relay case over  the integrated RIS and relay case is reduced upon increasing $y^{\text{RIS}}$. More specifically, the best positioning for the RIS is near of the S for the joint RIS and relay transmission, while its proximity to D is the best option for the integrated RIS and relay scheme.

\begin{figure}[!t]
	\begin{center}
		\includegraphics[width=0.9\columnwidth]{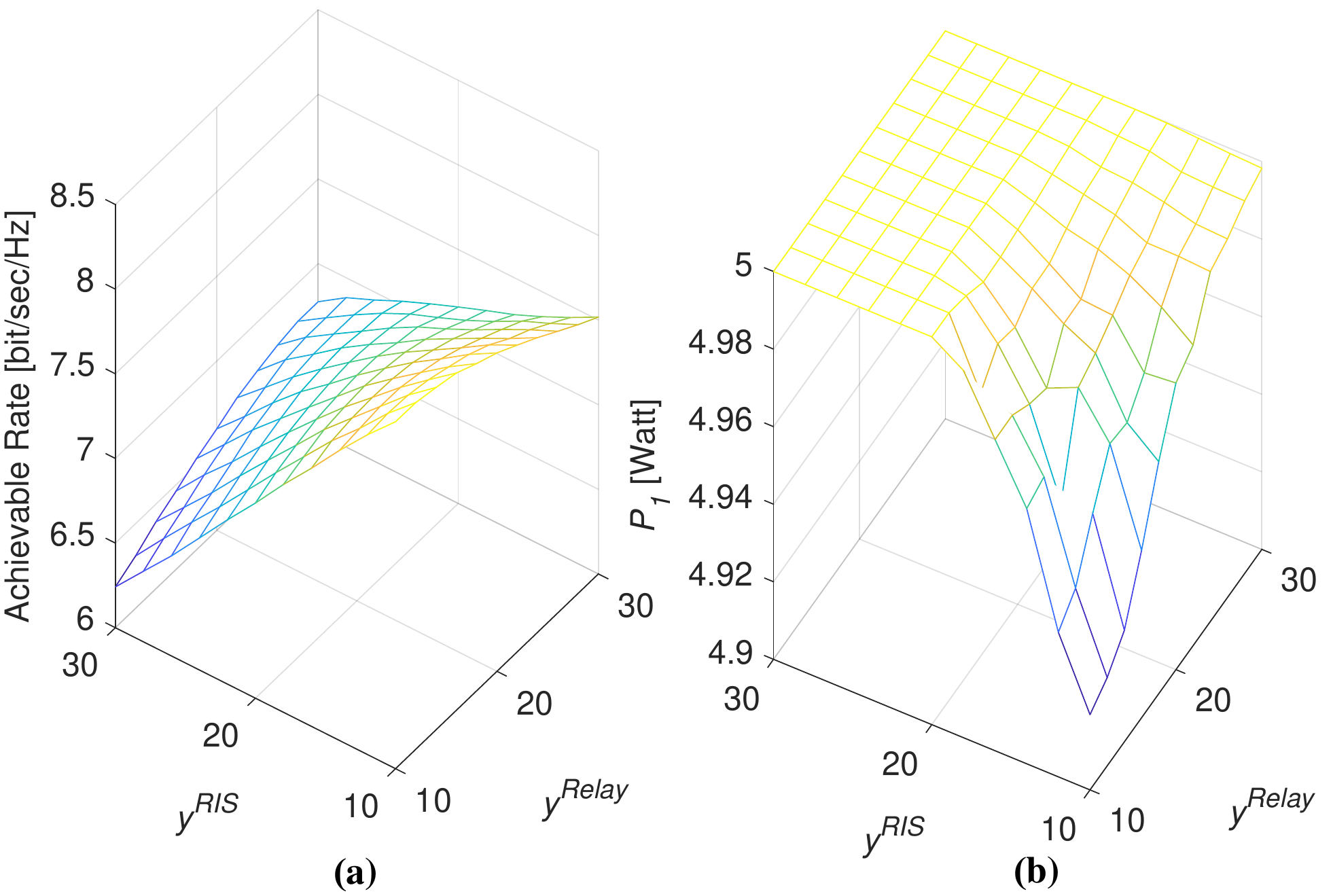}
		\vspace*{-0.3cm}\caption{(a) Achievable rate and (b) $P_1$ analysis for the joint RIS and relay transmission with the sequential algorithm for $N=1024$ at 28 GHz.}\vspace*{-0.6cm}
		\label{fig:Opt}
	\end{center}
\end{figure}

In Fig. \ref{fig:Opt}(a), the achievable rate of the joint RIS and relay transmission scheme for varying RIS and relay positions in the $y$-axis is analyzed by considering the proposed sequential optimization algorithm under the non-ideal relay assumption at $ 28$ GHz. In Fig \ref{fig:Opt}(b), the $ P_1 $ values that maximize the achievable rate are plotted for varying RIS and relay locations in the $y$-axis. Here, $P_\text{tot}=5$ W and the S,  D,  RIS, and the relay are respectively located at $(10,0)$, $(10,40)$, $(0,y^{\text{RIS}})$ and $(0,y^{\text{Relay}})$ in $xy$-plane.
As shown from Fig. \ref{fig:Opt}(a), the achievable rate is maximized when both the RIS and the relay are positioned close to  S. If the RIS is positioned far from  S, the quality of the signals received by the relay will be deteriorated considerably. Since the perfect decoding assumption is not considered, the signal quality at the relay increases as the relay is located closer to the RIS, therefore, the required $ P_1 $ decreases, and more power is assigned to the second-time slot, as seen from Fig. \ref{fig:Opt}(b). In other words, the capacity is maximized when the relay is activated and more power assigned to $ P_2 $.

\vspace*{-0.2cm}

%
%In Figs. (a) and (b), the effect of varying Rx positions through the $ x $ and $ y $-axis on the achievable rate of an RIS-assisted system at $28$ GHz is examined for a UMi Street Canyon outdoor environment and $N=256$. 
%Here, we consider Scenario 1 where the RIS is mounted on the $xz$ plane and the coordinates of the Tx, the Rx, and the RIS are respectively given as $(0,25,20)$, $(x^{\text{Rx}},y^{\text{Rx}},1)$, $(70,85,10)$. In Fig. \ref{fig:Sim3}(a), the direct link between the Tx-Rx is available as well as the RIS-assisted link for transmission, while it is assumed that the direct link between Tx-Rx is blocked in Fig. \ref{fig:Sim3}(b). In Fig. \ref{fig:Sim3}(a), we observe that the highest achievable rate is obtained when the Rx is close to the Tx, since the channel between Tx-Rx is more dominant than the RIS-assisted link in terms of achievable rate. Furthermore, if the Rx moves to an area far from the RIS on the $ x $ and $y$-axis, the effect of the RIS will drastically diminish as shown in Fig. \ref{fig:Sim3}(b). We observe that the most decisive performance parameter is the separation between the RIS-Rx when the direct link is blocked between the Tx-Rx.
%

\vspace*{-0.2cm}
\section{Conclusions}
In this letter, we have proposed two hybrid transmission concepts  to improve achievable rate and error performance by exploiting the most important advantages of RIS and relaying technologies. Although, most of the previous works consider RISs and relaying as rivals to each other,  we have shown by extensive numerical studies and theoretical derivations that these two exciting technologies can work in a harmony with each other under appropriate transmission scenarios. It is also demonstrated through simulations that the relay can be considered as an additional performance-enhancing component to the RIS-assisted transmission scenario especially in the case of rapid deterioration in the channel.
 \vspace*{-0.3cm}

\bibliographystyle{IEEEtran}
\bibliography{bib_2020}

\end{document}